# Measurement of Neutrino Source Activity in the Experiment BEST by Calorimetric Method

**V.N. Gavrin[a], T.V. Ibragimova[a], J.P. Kozlova[a,*], V.A. Tarasov[b], E.P. Veretenkin[a] and A.I. Zvir[b]**

[a] *Institute for Nuclear Research of the Russian Academy of Sciences, Prospect 60-letiya Oktyabrya, 7a, Moscow, 117312, Russian Federation*

[b] *Research Institute of Atomic Reactors, Zapadnoye Shosse, 9, Ulyanovsk region, Dimitrovgrad, 433510, Russian Federation*

*E-mail:* julia@inr.ru

ABSTRACT: Experiments to study non-standard properties of neutrinos using high-intensity artificial neutrino sources require high accuracy in determining their activity. A calorimetric system to determine the activity of a $^{51}$Cr neutrino source in the experiment BEST with an accuracy better than 1% was created. The design of the calorimetric system and the main factors affecting the accuracy of measurements are considered. The results of measurements of the activity of the source made for the BEST experiment are presented. The value of the source activity from calorimetric measurements was 3.41 MCi at the beginning of the experiment.

[*]Corresponding Author

# Contents

## 1. Introduction

Currently the experiment BEST [1] is being conducted in the Baksan Neutrino Observatory (BNO) INR RAS. The purpose of this experiment is to study the possibility of oscillations of electron neutrinos into sterile states with $\Delta m^2 > 0.5$ eV$^2$ and $\sin^2 2\theta > 0.1$. The idea of the experiment is to irradiate two gallium targets located at different distances from the artificial high-activity neutrino source. A two-zone gallium target was created in the underground laboratory of the gallium-germanium neutrino telescope (GGNT) (figure 1).

The detector consists of an external cylindrical and an internal spherical container. In the center of the spherical container, a channel is provided for placing the neutrino source. The internal (first) gallium target in the form of molten metal Ga with a mass of 7.40 t is in the spherical container in the immediate vicinity of the source. The average neutrino path in this target is 53 cm. In the gap between the spherical and cylindrical containers, there is an external (second) gallium target with a mass of 40.09 t Ga. The average neutrino path in the external target is 55 cm.

By comparing the neutrino capture rates in the inner and outer zones, as well as the expected and measured capture rates, we plan to obtain evidence for the existence or absence of short-baseline neutrino oscillations. The neutrino capture rate is determined using the scheme worked out for a gallium radiochemical detector, which includes the chemical extraction of $^{71}$Ge formed because of the interaction of the neutrino with the gallium target, and the determination of the amount of $^{71}$Ge in the proportional counter [2].

For the BEST experiment, an artificial neutrino source based on $^{51}$Cr (the half-life is $T_{1/2} = 27.703 \pm 0.003$ days [3] with activity > 3 MCi was manufactured at the Research Institute of Atomic Reactors (RIAR, Dimitrovgrad) (figure 2).

The source is a set of 26 chromium disks with diameters of 84 and 88 mm, a thickness of 4 mm and a total weight of 4007.5 g. The chromium disks made of metallic $^{50}$Cr enriched to 97% were irradiated for ~ 100 days with thermal neutrons in the SM-3 reactor to produce $^{51}$Cr. After irradiation, the disks were placed in a hermetic stainless-steel capsule, a biological protection shield based on a tungsten alloy with a thickness of ~ 30 mm and a weight of 42.8 kg, and a steel shell with a special cap for capturing the source by a manipulator.

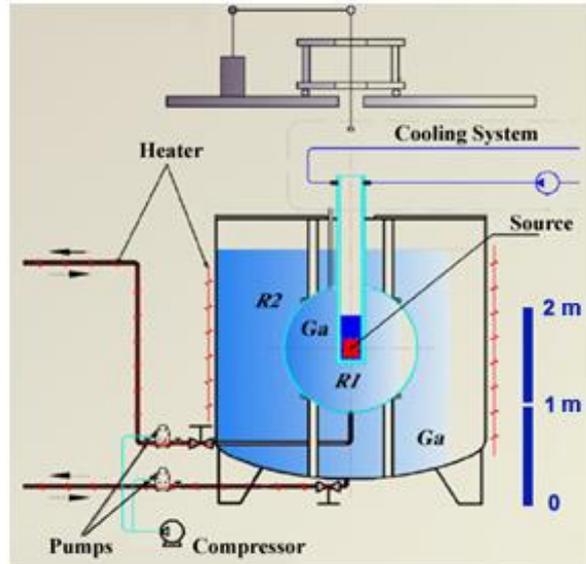

**Figure 1.** Two-zone gallium target.

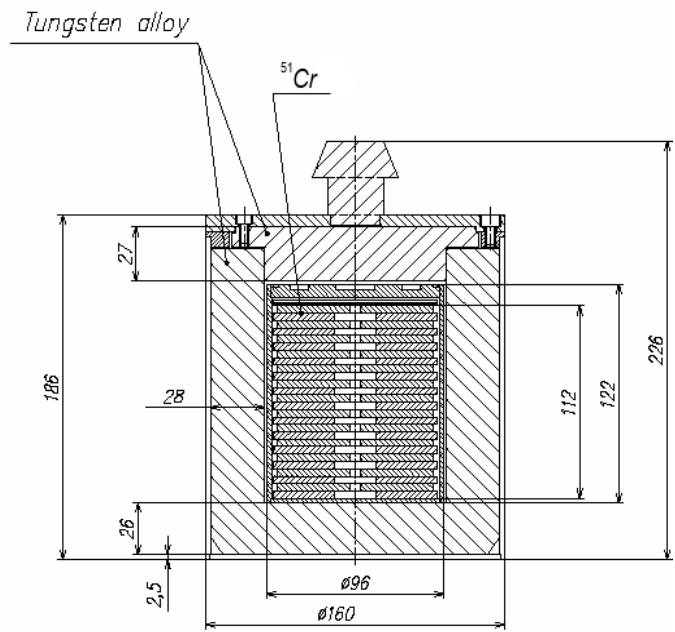

**Figure 2.** Neutrino source based on $^{51}$Cr, all dimensions are in mm.

One technique to determine the activity A of a neutrino source is a calorimetric method based on the measurement of heat release N from $^{51}$Cr decay. $^{51}$Cr decays by electron capture to the

– 2 –

ground state of the $^{51}$V (90%) and to the excited state of the $^{51}$V$_m$ (10%), which then decays to the ground state, emitting a 320 keV gamma ray. The gamma-ray absorption makes the main contribution into the source heat release that can be measured by calorimetric system [4]. The activity of the source is defined as $A = \frac{N}{\varepsilon}$, where ε is the average energy release per $^{51}$Cr decay 36.750 ± 0.084 keV/decay as shown in Reference [4]. It is important to notice that the uncertainty on ε of 0.084 keV/decay is the main systematic uncertainty on the power-to-activity conversion. Using the ε value and the SI system values of the eV and the Ci the conversion factor f of 217.856 ± 0.498 W/MCi is obtained, with which we calculate $A = \frac{N}{f}$ for each experimental run. This article describes the design of a calorimetric system created to measure the activity of the specified source, the calibration of this calorimetric system using a heat simulator and the measurement results obtained during the BEST experiment.

## 2. Calorimetric system

### 2.1 Principle of calorimetric method

The choice of method and type of calorimeter is determined by the nature and duration of the measurement process, the temperature range in which the measurements are made, the amount of heat to be measured, and the required accuracy. A differential calorimeter was used in the SAGE experiment to measure the activity of neutrino sources based on $^{51}$Cr and $^{37}$Ar [5, 6]. An isothermal water-bath calorimeter was used to assay radioactive samples with thermal output up to 1000 W [7]. A calorimeter of the same type was proposed to measure the activity of the $^{144}$Ce-$^{144}$Pr antineutrino source in the SOX experiment [8].

In accordance with the requirements of the BEST experiment, the calorimetric system created must meet the following basic conditions:

- the range of heat release measurements is 50-740 W, which corresponds to the source activity of 0.2-3.4 MCi,
- the time required for measurement is no more than 20 hours,
- the total error in measuring source activity is no more than 1%.

Due to the need to measure the heat release of hundreds of watts, the flow-mass calorimeter was chosen (figure 3), which provides complete removal of a significant amount of heat.

The principle of measuring heat release by a flow-type calorimeter is a transfer of all heat from source to coolant. In this case, at a constant flow rate, the difference in the specific enthalpy of the coolant at the outlet and at the entrance to the calorimeter is directly proportional to the heat release:

$$N = Q\big(H(p, T_{out}) - H(p, T_{in})\big) \quad (1)$$

where N is the heat release of the source, W,
Q is the coolant flow rate, kg/s,
H – the specific enthalpy of the coolant as a function of the temperature and pressure, J/kg,
T$_{out}$ is the coolant temperature at the outlet of the heat exchanger, °C,
T$_{in}$ is the coolant temperature at the entrance to the heat exchanger, °C.

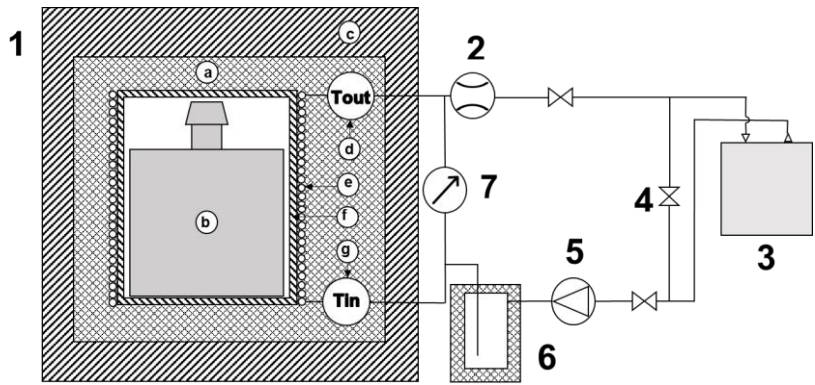

**Figure 3**. Schematic hydraulic diagram of the mass flow calorimeter: 1 – measuring cell: a-thermal insulation, b-neutrino source, c-biological protection, d-output thermal resistance, e-heat exchanger, f-container, g-input thermal resistance; 2 – flow meter; 3 – thermostat with a circulation pump; 4 – bypass; 5 – gear pump, 6 – temperature buffer, 7 – differential manometer.

**2.2 Experimental equipment**

For measurements, the source (b) is placed in the container (f) of the measuring cell 1 (figure 3). The container is surrounded by a labyrinth heat exchanger (e). To prevent heat losses, the container with the heat exchanger is surrounded by thermal insulation (a), which, in turn, is surrounded by biological protection (c), which is necessary when working with a radioactive source.

The input ($T_{in}$) and output ($T_{out}$) temperatures of the coolant are measured by platinum thermal resistances PTV 2-1 (g) and (d), respectively, with an uncertainty of no more than 0.002°C. The cooling thermostat Unistat 405 (3) maintains a constant temperature at the inlet of the heat exchanger with a temperature stabilization accuracy of 0.1 °C. For the early measurements of the source heat release, these values were: $T_{in}$ ~ 12 - 15 °C, $T_{out}$ ~ 26 - 23 °C. The constant flow of the coolant through the heat exchanger (~ 40 kg/h) is provided by the precision gear pump Ismatec Reglo-Z Digital (5) and is measured by the Coriolis flow meter Micro Motion CMF010 (2) with an uncertainty of ±0.05%. The bypass line (4) and the temperature buffer (6) contribute to more accurate temperature control. The Delta-plus DPG40 differential manometer (M) controls the differential pressure of the coolant on the measuring cell. Deionized water is used as the coolant.

**2.2.1 Simulation of thermal characteristics of the apparatus**

An appropriate simulator was manufactured to calibrate the calorimeter (figure 4). The simulator is a duralumin block with 4 electric heaters in the central part. For comparison, the dash-dotted line shows the zone where heat is generated in a real neutrino source. The duralumin block is placed in a stainless-steel shell, the geometric dimensions of which correspond to the size of the neutrino source. The total heat capacity of the simulator and the source were estimated from the expression $C = \sum_i x_i c_{pi}$, where $x_i$ and $c_{pi}$ are the mass and specific heat of the constituent substances with an error of ~2%. The heat capacity of the heat simulator was 10030 J/K, the heat capacity of the source was 9860 J/K. These values were close to each other. The thermal conductivity of duralumin is close to the thermal conductivity of tungsten alloy of the biological protection of the source (160 W/m·K and 163 W/m·K, respectively). A Sorensen XFR 300-3.5 precision DC power source was used to heat the simulator. The supplied electrical power was determined from measurements of the voltage drop on the heaters and the measuring shunt according to the readings of a 10-channel universal voltmeter B7-78/1. The error of measuring electrical power on the simulator is 0.04 %, the main contribution to which is the uncertainty of the resistance of the measuring shunt.

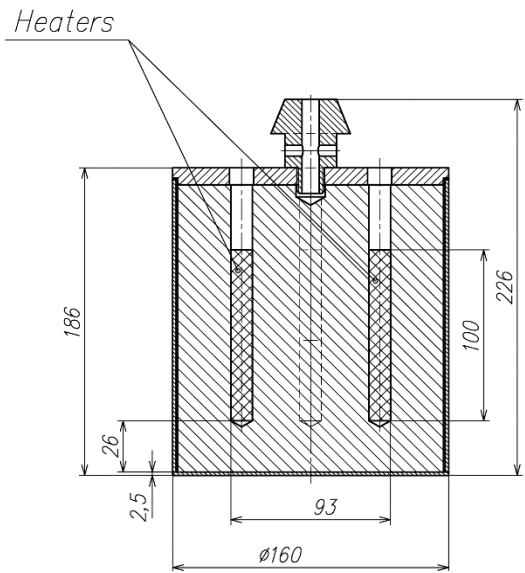

**Figure 4**. The source heat simulator, all dimensions are in mm.

**2.2.2 Operation of the experiment**

To control the operation of the main devices of the calorimetric system (flow meter, temperature meter, voltmeter, and power supply), software was developed based on the National Instruments LabVIEW package. The control program allows reading and saving the main and auxiliary parameters of the system for several hours of measurement, calculates the activity of the source taking into account heat loss to the environment and hydrodynamic friction in the heat exchanger, and also provides electrical power to the source heat simulator in static and dynamic modes. To ensure radiation safety, the calorimetric system is controlled remotely. Origin Pro was used for plotting and numerical data analysis, including linear and nonlinear curve fitting (version OriginPro 2019b 64bit).

To measure the heat release of the source in the time allotted for measurements, no more than 20 hours, it is necessary that the calorimetric system reaches thermal equilibrium quickly enough. Figure 5 shows the time dependence of the input and output temperatures of the coolant and the sidewall of the container of the measuring cell after applying a fixed power of 500 W to the simulator.

The thermal time constant of the calorimeter, determined from these dependencies, is ~ 50 min. The thermal equilibrium of the system is established after 6 hours of measurements, when the fluctuation of all temperature values is ± 0.002 K. Therefore, the time interval of 10-20 hours is sufficient for a proper measuring of the source heat release.

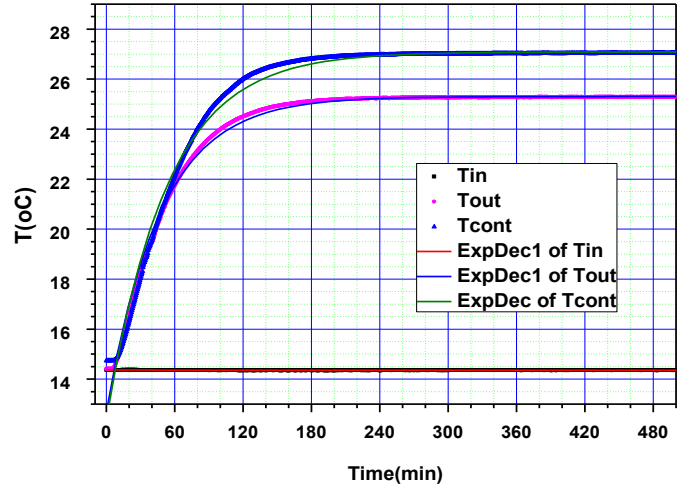

**Figure 5**. Time dependence of the input and output temperatures ($T_{in}$ and $T_{out}$) of the heat carrier and the sidewall of the container of the measuring cell ($T_{cont}$) when the heat simulator is supplied with a power of 500 W. The red line is the result of approximating the temperature dependences by an exponential relationship: $T = Ae^{\left(-\frac{t}{b}\right)} + T_0$.

### 2.2.3 Calorimeter calibration

The calibration dependence of the calorimeter at thermal equilibrium was studied. It was the dependence of the temperature difference of the coolant at the output and input of the calorimeter on the heat release corresponding to the different electrical power supplied to the thermal simulator $P_{set}$. The system was calibrated using the thermal simulator with constant electrical power $P_{set}$ in the range 0 – 610 W, as well as before each measurement of the source activity with $P_{set}$ close to the expected heat release of the source within 1%. In addition, the calorimetric system was calibrated without supplying electrical power to the heat simulator, and the $\Delta T = T_{out} - T_{in}$ was determined in the absence of heat loss to the environment.

Statistical processing of the data array was performed in the time interval of ~ 6-9 hours of system operation under steady-state thermal equilibrium. After 10 hours of operation, the dynamic mode was set, simulating the radioactive decay of $^{51}$Cr with a half-life of 27.703 days [3] to account for the delay in changing when changing $P_{set}$.

As an example, figure 6a shows the time dependence of the input and output temperatures of the coolant when the initial electrical power $P_{set}$ = 162.5 W is applied to the simulator, as well as the temperatures of the container wall and the environment. Figure 6b shows the corresponding time dependence ΔT, calculated in the control program from measurements of the output and input temperatures of the coolant.

Figure 7 shows the relation of the constant electrical power *$P_{set}$* supplied to the simulator and the difference of output and input temperatures of the coolant. The heat release was normalized to the ratio of the measured value of the coolant flow Q to the constant value of 40kg/h: $P_{hnorm} = P_{set}\frac{40}{Q}$.

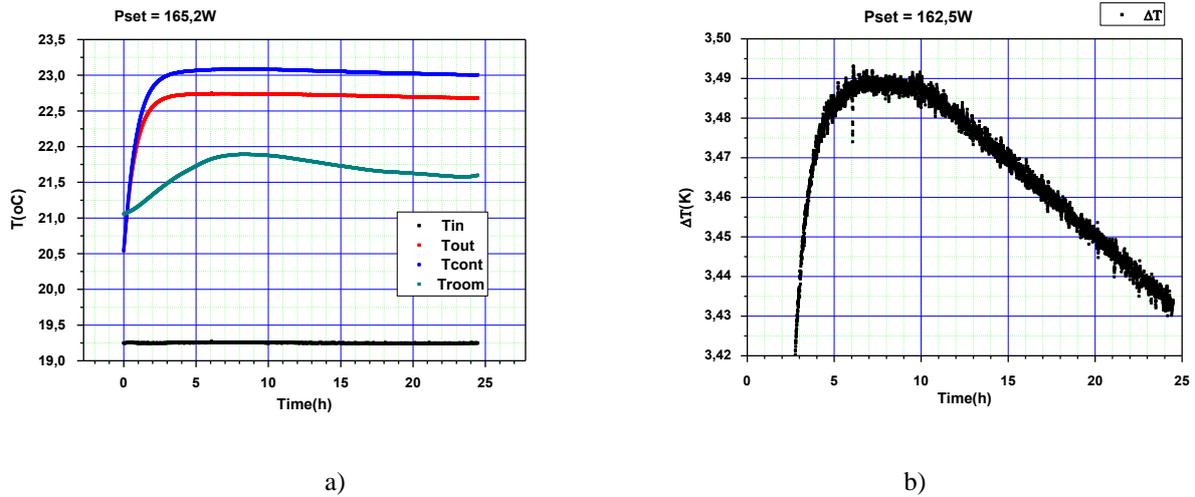

a)                                                                              b)

**Figure 6**. a) The time dependence of the temperatures of the coolant, container wall and environment when the initial electrical power of 162.5 W is applied to the simulator with a subsequent exponential decline, b) the corresponding time dependence ΔT.

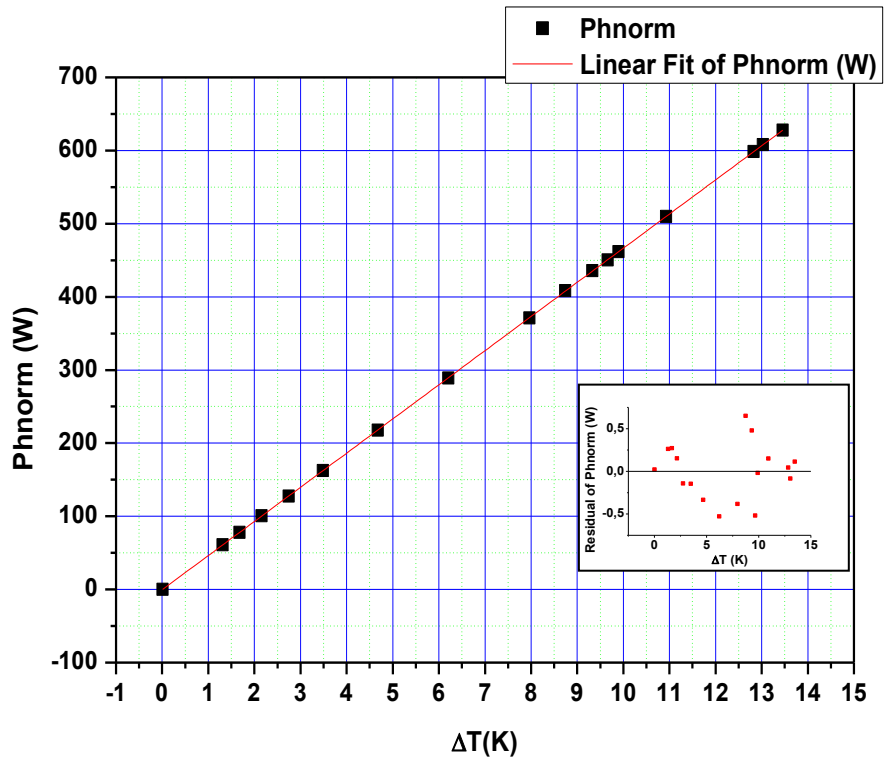

**Figure 7**. Calibration dependence of the difference between the input and output temperatures of the coolant on the value of normalized heat release. ■ - measured points, solid line - linear approximation, ■ – residuals of normalized heat release (on the insert).

The obtained calibration values were approximated by a polynomial of the first degree: $P_{hnorm} = (46.689 \pm 0.019) \cdot \Delta T - (0.41 \pm 0.15)$, with $\chi^2/\text{DOF}(15) = 0.1$. In the absence of heat loss, considering the heat capacity of deionized water, which is 4184.1 J/kg·K [9], the proportionality coefficient should be equal to 46.49 W/K. The difference between the theoretical and experimental coefficient obtained from the data approximation indicates the presence of heat loss to the environment of the order of 0.4% during calibration.

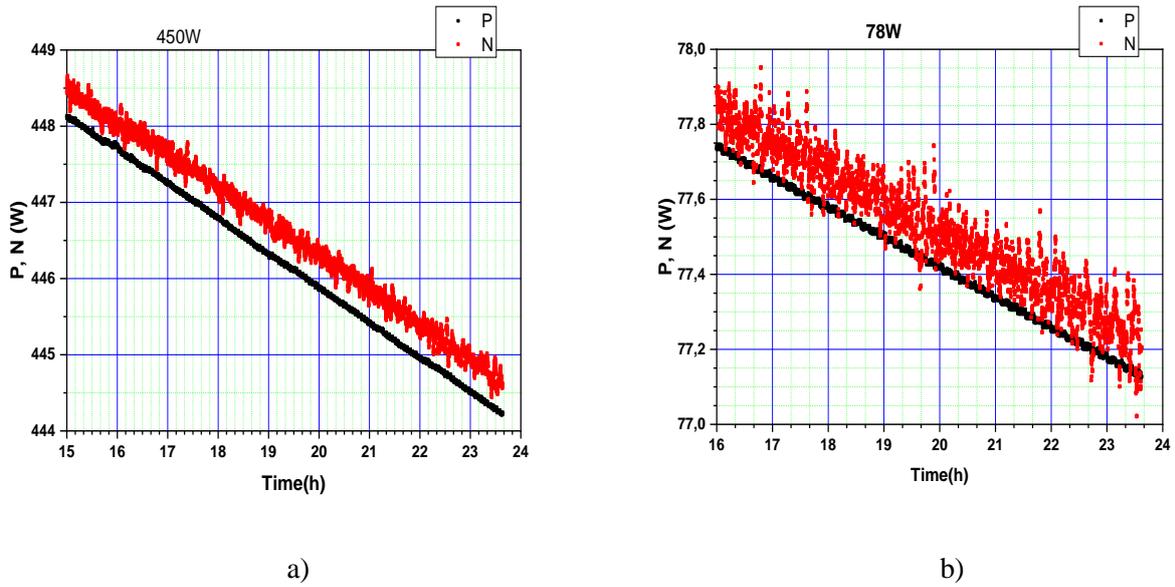

a)                                                                         b)

**Figure 8**. Time dependence of heat release of the thermal simulator in the dynamic mode at initial heat capacity 450 W (a), and 78 W (b); P (black) - the actual heat release in the simulator, N (red) – heat release determined by the calorimeter.

To study the effect of the decay of $^{51}$Cr on the measurement results, the calorimeter was calibrated by the thermal simulator in dynamic mode, when the heat release of the simulator changed exponentially in accordance with the decay of $^{51}$Cr, the half-life of which is $T_{1/2} = 27.703 \pm 0.003$ days [3]. The measurements were made at different values of initial heat release on the simulator. Figure 8 shows examples of the results of these measurements at initial heat releases of 450 W (a) and 78 W (b). When the power supplied decreases, the excess heat is carried away by the coolant for some time, that is, there is a delay in the change in ΔT relative to the change in $P_{set}$. In the heat release range of 580-60 W, the delay of the calorimeter measurements N from the actual heat release value P is 40-90 minutes, that was taken into account in the measurements of the neutrino source heat release.

The calorimeter calibrations to determine the dependence in the entire heat release range of 50-700 W were not carried out under identical conditions. For example, to reduce heat loss to the environment it was necessary to increase the input temperature of the coolant for the decreasing heat release of the thermal simulator and, accordingly, its flow rate changed.

Therefore, to reduce the influence of external factors on the measurement of heat release, before each measurement with a source, the calorimeter was calibrated at the heat release that differed from the expected heat release of the source by no more than 1%. The exception is the first measurement in which calibration was performed after the source activity was measured, since only the estimated activity value was known before the first measurement. The values of the input temperature and coolant flow were maintained the same in each specific measurement with the simulator and the source. In addition, the calorimetric system was calibrated without supplying electrical power to the heat simulator, and the ΔT value was determined in the absence of heat loss to the environment. Linear dependencies $N = k_i Q(T_{out} - T_{in})$ were obtained from two calibration points for each specific measurement of heat release and were used for source activity measurement.

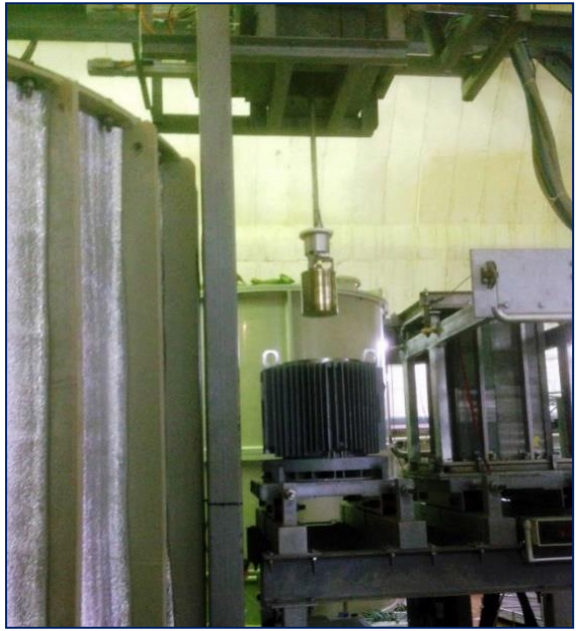

**Figure 9**. Part of the BEST installation: the neutrino source above the transport container (in the center), a container with gallium targets (on the left), a measuring cell of the calorimeter system (on the right).

### 3. Analysis and results of the calorimetric measurements

The artificial neutrino source was delivered to the BNO on July 05, 2019 and installed in the channel of the two-zone gallium target of the BEST installation at 14:02.

Figure 9 shows the moment when the source was moved from the transport container to the channel of the gallium target. After nine days of irradiation of gallium targets with a neutrino stream from the source, the resulting $^{71}$Ge was extracted from the gallium targets. At that time, measurements of photon spectra from the source and identification of lines from impurity radioactive elements in the source were performed to assess the contribution of their heat release to the total heat release of the source. The photon spectra from the source and a list of detected radioactive impurities, their activity and heat release are presented in figure 10 and table 1 [10].

The contribution of radioactive impurities (2.9 mW) to the total heat release of the source was not considered since it is about $5 \cdot 10^{-6}$ in the total heat release of the $^{51}$Cr source (~740 W).

After spectrometric measurements, the source was moved to the calorimetric system to determine the activity by its heat release. During the measurements, all important parameters were measured and recorded: the coolant temperature at the inlet and outlet of the measuring cell, the temperature on the side wall of the container of the measuring cell under thermal insulation, and the ambient temperature. Temperature measurements were made at intervals of 25 seconds. The coolant flow rate (Q) was measured and recorded every second. In total, ten cycles of calorimetric measurements of the source heat release were performed with an interval of 9-10 days between measurements.

The source activity was determined by its heat release in three ways. The results are consistent and this procedure provides protection against unforeseen systematic errors. The method that lead to the best agreement with the known half life for $^{51}$Cr was used for the final result. For the first method, the activity was calculated directly using the enthalpy change in the coolant. The electric calibration system was used to determine the heat loss correction. For the second method, the heat from the source is determined using the electrical calibration data for the calorimeter response. For the third method, the electric calibrator was used but in this case the response of the calorimeter was determined using the calibrator at the actual conditions of the source measurement rather than using the mean calibration coefficient. The methods are detailed below.

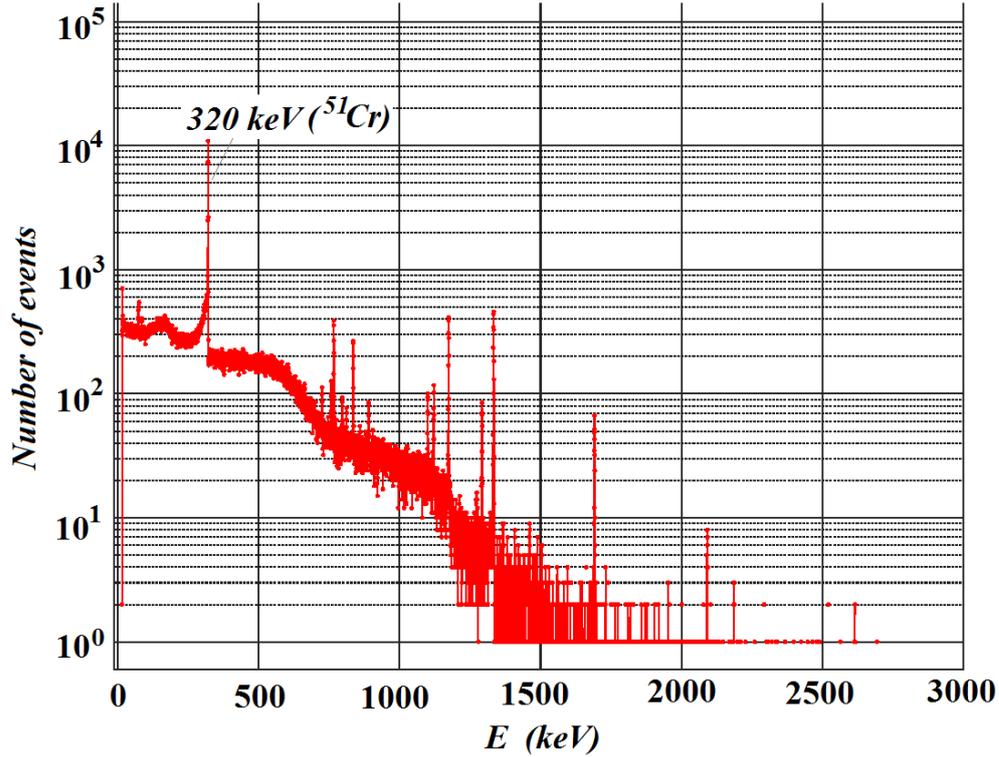

**Figure 10.** Full photon spectra from the $^{51}$Cr source [10].

**Calculating method.** To calculate the source activity the expression (2) was entered in the control program and considered heat loss and the contribution of hydrodynamic friction in the measuring cell of the calorimetric system:

$$A_{calc} = \frac{[N_{heat} + N_{loss} - N_{resist}]}{f} =$$
$$[Q(H(p, T_{out}) - H(p, T_{in})) + l_i(T_{cont} - T_{room}) - \Delta p \cdot Q/\rho]/f \quad (2)$$

where $A_{calc}$ – the source activity, MCi; $N_{heat}$ – the transferring coolant heat, W; $N_{loss}$ – the heat losses, W; $N_{resist}$ – the heat release due to hydraulic resistance in heat exchanger, W; f – the scaling factor, 217.856 W/MCi; Q – the coolant flow rate, kg/s; H – the specific enthalpy of the coolant as a function of the temperature and pressure, J/kg; $l_i$ – the heat loss coefficient (obtained from i-calibration experiment), W/K; $T_{cont}$ – the temperature of container side, °C; $T_{room}$ – the room temperature, °C; Δp – the differential pressure in heat exchanger, Pa; ρ – the coolant density, kg/m$^3$.

**Table 1.** Measured radionuclide impurities in the $^{51}$Cr neutrino source and their contribution to the heat release of the source on July 5, 2019 [10].

| | Isotope, $T_{1/2}$ | Energy, keV | Line output, % | Activity (05.07.2019), mCi | Heat release, mW |
|---|---|---|---|---|---|
| 1 | $^{137}$Cs, 30.05 y | 662 | 85 | 8.5·(1±0.23) | 0.06 |
| 2 | $^{95}$Zr, 64 d | 724<br>757 | 11.1<br>54.38 | 60·(1±0.12) | 2.1 |
| 3 | $^{95}$Nb, 35 d | 766 | 99.8 | 87·(1±0.04) | |
| 4 | $^{134}$Cs, 2.06 y | 796 | 85.5 | 3.3·(1±0.18) | 0.041 |
| 5 | $^{58}$Co, 70.85 d | 811 | 99.44 | 6.0·(1±0.27) | 0.08 |
| 6 | $^{54}$Mn, 312 d | 835 | 100 | 13·(1±0.05) | 0.10 |
| 7 | $^{46}$Sc, 83.8 d | 889<br>1120 | 100<br>100 | 5.2·(1±0.10) | 0.07 |
| 8 | $^{59}$Fe, 44.5 d | 1099<br>1291 | 57<br>43.2 | 23·(1±0.07) | 0.22 |
| 9 | $^{60}$Co, 5.27 y | 1173<br>1332 | 100<br>100 | 6.6·(1±0.03) | 0.11 |
| 10 | $^{124}$Sb, 60.2 d | 1690<br>2091 | 47.5<br>5.5 | 5.8·(1±0.06) | 0.10 |
| 11 | $^{154}$Eu (?), 8.6 y | 1274<br>1595 | 34.9<br>5.5 | 0.86·(1±0.18) | 0.10 |
| Σ | | | | | 2.9 |

The contribution of hydraulic friction was (0.4 ± 0.1) W, which is less than 0.7 % of the source heat release. The contribution of heat loss to the environment, estimated in calibration experiments, was 2.5 to 0.5 W in the range of heat release 600-60 W, which is less than 1% of the source heat release.

Figure 11 a) shows an example of measuring activity using this method. In the time interval of 0 – 10 hours, the thermal equilibrium of the hot source with the calorimetric system was established. After 10 hours of measurements, the source activity decreased in accordance with the decay of $^{51}$Cr (figure 11b)).

In the column $A_{calc}$ (table 2) the results of determining the source activity using the method given in expression (2) for each of the 10 measurements performed are presented. Experimental data A are approximated by the dependence $A_{calc} = a \cdot e^{-bt}$ with a fixed value of the coefficient b corresponding to $T_{1/2}$ = 27.703 days. $A_{calc}$ is assigned the value of the approximation coefficient a, reduced to absolute time, and considering the time delay. The reference activity $A_0$ for the time of installation of the source in a two-zone gallium target at 14:02 05.07.2019 is also given, obtained from the approximation of the time dependence of ten measured values of $A_{calc}$ by an exponent with a $^{51}$Cr half-life of $T_{1/2}$ = 27.703 days. The last row of table 2 shows the half-life value obtained from an exponential approximation of data with a variable parameter b.

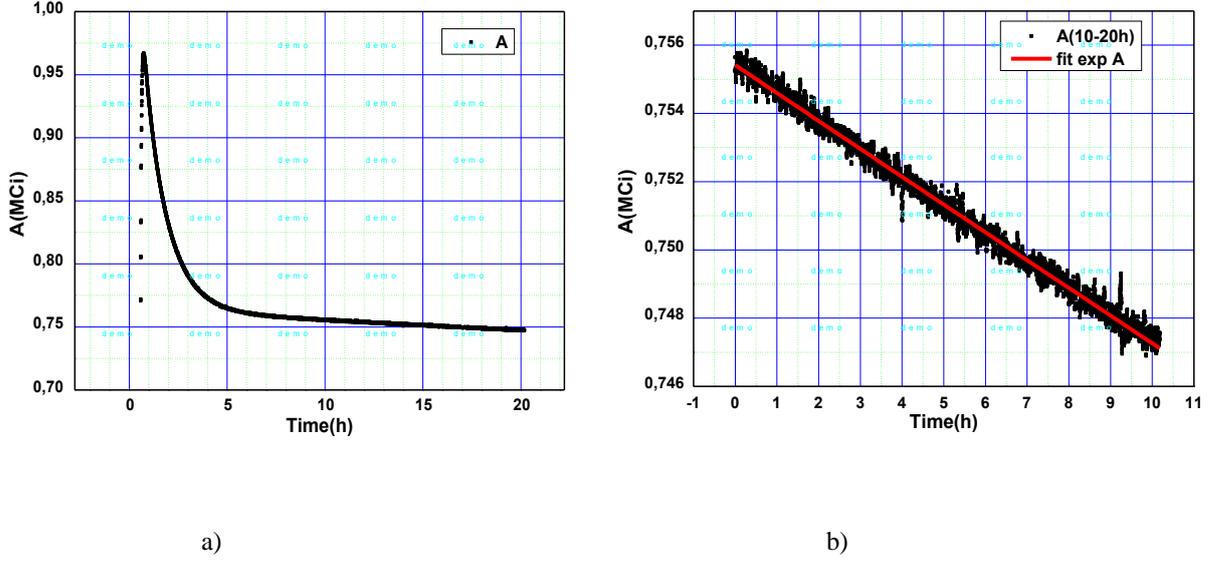

a) b)

**Figure 11.** Determination of the activity of the $^{51}$Cr neutrino source for the 6th calorimetric measurement (03.09.2019) by the calculated method a) in the full-time interval of measurements; b) in the measurement interval of 10-20 hours with an approximation by exponential decay $A_{calc} = a \cdot e^{-bt}$ with b fixed to a value corresponding to $T_{1/2} = 27.703$ days.

**Determination of the source activity by calibration dependence.** In this case, the calibration dependence of heat release (N) on the difference between the output and input temperatures of the coolant ($\Delta T$) was used, obtained from a linear approximation of calibration measurements with a thermal simulator (figure 7): $N = P_{hnorm} = (46.689 \pm 0.019) \cdot \Delta T - (0.41 \pm 0.15)$.

As an example, figure 12 shows the time dependence of $\Delta T$ obtained in measurements with the source (4th measurement 14.08.2019) and approximated by the dependence $\Delta T = a \cdot e^{-bt}$ with a fixed value of the coefficient b corresponding to $T_{1/2} = 27.703$ days. According to these data, $\Delta T$ was determined at the fifteenth hour of measurements as the average value in the range of 14.5 – 15.5 hours. The average value in the interval of 14.5 – 15.5 hours was $\Delta T = 6.1254 \pm 0.0019$ K. The value of $\Delta T$ at 15 hour obtained from the approximation was 6.1256 K and, taking into account the uncertainty of the temperature measurement (0.002 K), did not differ from the average value in the time interval of 14.5-15.5 hours. For each point, the value of $\Delta T$ was corrected for the actual coolant flow at the time of measurement using $\Delta T_{norm} = \Delta T * \frac{40}{Q}$, where Q is the flow at the measuring time. The nominal flow was 40 kg/h.

The results of 10 measurements of activity ($A_{cal} = N/f$), $A_0$ activity on 14:02 05.07.2019 and $T_{1/2}$ obtained by this method are presented in the $A_{cal}$ column of table 2.

**Determination of the source activity from two-point calibration.** The activities $A_{heat} = N_i/f$ were calculated from linear dependencies $N = k_i Q(T_{out} - T_{in})$ which were obtained using two calibration points: at the simulator heat release that was close to the expected heat release of the source and at zero electric power on the simulator. The $\Delta T$ values corresponded to the $\Delta T$ values in the calibration method. Figure 13 demonstrates the time dependence of $A_{heat}$, calculated by this method; the blue dash line is an exponential decay approximation $A = a \cdot e^{-bt}$ with a fixed half-life ($T_{1/2}$ = 27.703 days), the red line is an approximation with coefficient *b* from experimental data. The $A_{heat}$ column of table 2 shows $A_{heat}$ values at different data, $A_0$ activity on 14:02 05.07.2019 and calculated $T_{1/2}$ obtained by this method.

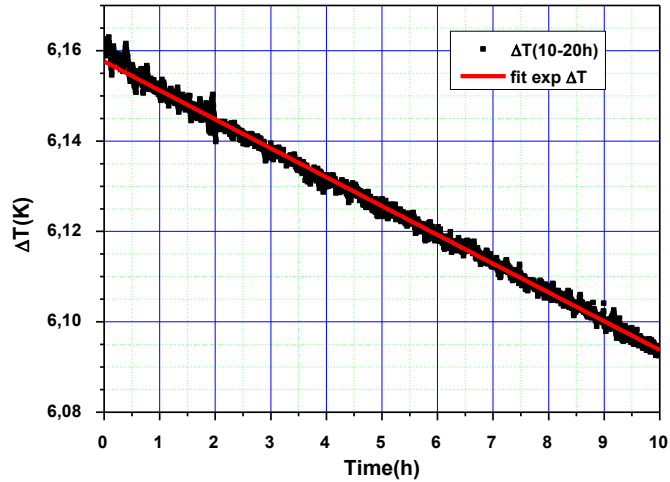

**Figure 12**. The time dependence of the temperature difference ΔT of the coolant obtained in the 4th measurement of the source activity (14.08.2019) with an approximation by exponential decay $\Delta T = a \cdot e^{-bt}$ with b fixed to a value corresponding to $T_{1/2}$ = 27.703 days.

All three methods of the source activity calculation are in a good agreement and give both values and errors close to one another. However, the standard error of the $A_0$ value obtained from the exponential approximation of 10 $A_{cal}$ measurements calculated by calibration method has minimal value 0.0010 MCi (0.029%), the $\chi^2$ value obtained by this method is minimal and the half-life of $^{51}Cr$ = 27.710±0.017 days is closed to the reference value of 27.703 ± 0.003 days [3]. Summary of the systematic uncertainties associated with the source activity are shown in table 3.

## 4. Conclusions

Adding the statistical and systematic uncertainties in quadrature gives the final value of the activity (A) of the artificial neutrino source $^{51}Cr$ in the BEST experiment at the time of installation of the source in a two-zone target at 14:02 05.07. 2019

$$A = 3.414 \pm 0.008 \text{ MCi}.$$

## 5. Acknowledgments

The work was performed using equipment of the unique scientific installation GGNT BNO INR RAS with the financial support of the Ministry of education and science of the Russian Federation: agreement No. 14.619.21.0009, unique project ID RFMEFI61917X0009.

The authors thank the State Atomic Energy Corporation "Rosatom" for the comprehensive support and fruitful cooperation in the implementation of the BEST experiment. The authors are also grateful to H. Robertson and D. Sinclair for useful discussions and an assistance in preparing the article for publication.

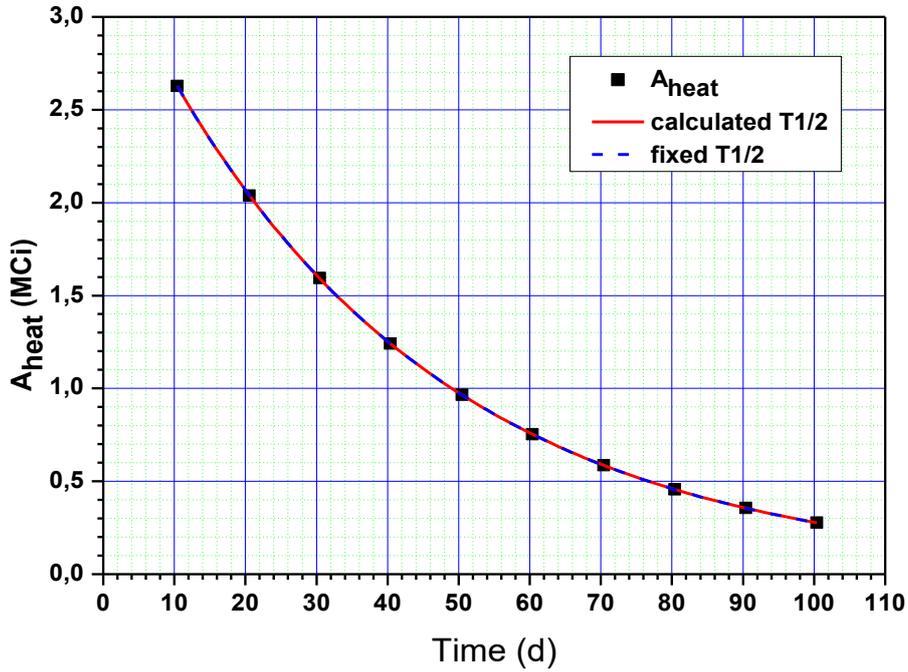

**Figure 13.** The activity of the neutrino source obtained by the two-point calculated method.

**Table 2**. The activity of the neutrino source in the experiment BEST obtained by three methods.

| № meas | Date, time | Days after 14:02 05.07.2019 | $A_{calc}$, MCi | $A_{cal}$, MCi | $A_{heat}$, MCi |
|---|---|---|---|---|---|
| 1 | 15.07.19 23-56 | 10.4125 | 2.624±0.020 | 2.6284±0.0134 | 2.6279±0.0085 |
| 2 | 26.07.19 03-23 | 20.5563 | 2.037±0.015 | 2.0404±0.0114 | 2.0381±0.0058 |
| 3 | 05.08.19 00-20 | 30.4292 | 1.591±0.010 | 1.5952±0.0080 | 1.5938±0.0039 |
| 4 | 14.08.19 23-16 | 40.3847 | 1.233±0.007 | 1.2442±0.0097 | 1.2419±0.0029 |
| 5 | 25.08.19 01-02 | 50.4583 | 0.962±0.007 | 0.9667±0.0072 | 0.9652±0.0023 |
| 6 | 03.09.19 23-05 | 60.3771 | 0.751±0.004 | 0.7539±0.0077 | 0.7532±0.0017 |
| 7 | 13.09.19 23-58 | 70.4139 | 0.584±0.003 | 0.5863±0.0070 | 0.5857±0.0013 |
| 8 | 23.09.19 23-45 | 80.4049 | 0.455±0.003 | 0.4557±0.0081 | 0.4556±0.0010 |
| 9 | 03.10.19 23-58 | 90.4139 | 0.354±0.0032 | 0.3546±0.0096 | 0.3559±0.0008 |
| 10 | 13.10.19 23-34 | 100.3972 | 0.276±0.002 | 0.2758±0.0065 | 0.2770±0.0006 |
| $A_0$, MCi | 05.07.19 14-02 | 0 | 3.4006±0.0019 | 3.4137±0.0010 | 3.4118±0.0011 |
| $\chi^2$ | | | 0.700125 | 0.1286969 | 1.6172998 |
| DoF | | | 9 | 9 | 9 |
| probability | | | 100% | 100% | 99.6% |
| $T_{1/2}$, d (Free variable) | | | 27.699±0.026 | 27.710±0.017 | 27.719±0.013 |

Table 3. Summary of the systematic uncertainties.

|   | Origin of uncertainty | Magnitude | Percentage |
|---|---|---|---|
| 1 | Power to activity conversion | 217.856 ± 0.498 W/MCi | 0.228 |
| 2 | $^{51}$Cr half-life | 27.703 ± 0.003 days | 0.011 |
| 3 | Constant flow of the coolant through the heat exchanger | 40.00 ± 0.02 kg/h | 0.05 |
| 4 | Resistance of the measuring shunt | 52.1513 ± 0.02 mOhm | 0.04 |
| 5 | Total uncertainty (added in quadrature) |  | 0.24 |